\documentclass[twocolumn,aps,pre,english,floatfix,superscriptaddress]{revtex4}
\pdfoutput=1
\usepackage{amsmath,amsfonts,amssymb,graphicx,bm,color,mathrsfs,verbatim}

\usepackage{soul}
\usepackage{ulem}
\usepackage{cancel}
\newcommand{\add}[1]{{\color{cyan}#1}}

\newcommand{\be}{\begin{equation}}
\newcommand{\ee}{\end{equation}}
\newcommand{\beq}{\begin{eqnarray}}
\newcommand{\eeq}{\end{eqnarray}}
\newcommand{\bml}{\begin{multline}}
\newcommand{\eml}{\end{multline}}

\def\la{\langle}
\def\ra{\rangle}

\begin{document}

\title{Negative mass corrections in a dissipative stochastic environment}
\author{Luca D'Alessio}
\affiliation{Department of Physics, The Pennsylvania State University, University Park, PA 16802, USA}
\affiliation{Department of Physics, Boston University, Boston, MA 02215, USA}
\author{Yariv Kafri}
\affiliation{Department of Physics, Technion, Haifa 32000, Israel}
\author{Anatoli Polkovnikov}
\affiliation{Department of Physics, Boston University, Boston, MA 02215, USA}

\begin{abstract}
We study the dynamics of a macroscopic object interacting with a dissipative stochastic environment using an adiabatic perturbation theory. The perturbation theory reproduces known expressions for the friction coefficient and, surprisingly, gives an additional \textit{negative} mass correction. The effect of the negative mass correction is illustrated by studying a harmonic oscillator interacting with a dissipative stochastic environment. While it is well known that the friction coefficient causes a reduction of the oscillation frequency we show that the negative mass correction can lead to its \textit{enhancement}. By studying an exactly solvable model of a magnet coupled to a spin environment evolving under standard non-conserving dynamics we show that the effect is present even beyond the validity of the adiabatic perturbation theory.
\end{abstract}
\maketitle

\section{Introduction}
In this article we analyze the dynamics of a macroscopic object coupled to a 
stochastic environment  which consists of many degrees of freedom. It is well 
known that coupling to an environment leads to dissipation which is typically 
described by an irreversible friction force. A classical example is the equation 
of motion describing a single macroscopic particle moving through a medium:
\begin{equation}
m\,\ddot{\boldsymbol{X}}+\eta\dot{\boldsymbol{X}}=-\frac{\partial V}{\partial\boldsymbol{X}}.\label{eq:langevin}
\end{equation}
Here $m$ and ${\bf X}$ are the mass and the position of the particle 
respectively, $\eta$ is the friction coefficient and $V$ includes both external
potentials and the interaction energy of the particle with the medium.  When the 
potential can be expanded around some equilibrium position ${\bf X}=0$, it becomes $V(\boldsymbol{X})\approx \frac{1}{2}m\,\omega_{0}^{2}\,\boldsymbol{X}^{2}$ and Eq.~\eqref{eq:langevin} 
describes a damped harmonic oscillator.  
In the under-damped regime $\zeta<1$,  
$\boldsymbol{X}$ oscillates with a \textit{shifted} frequency 
$\omega=\omega_{0}\sqrt{1-\zeta^{2}}$ which is always smaller than $\omega_{0}$. 
Here $\zeta=\eta/\left(2 m \omega_{0}\right)$ is the damping ratio. The 
amplitude of the oscillations decays to zero on a time scale $\omega_{0}\zeta$. 
When $\zeta>1$ the motion becomes over-damped and no oscillations are observed.

Equation~\eqref{eq:langevin} can be derived either using phenomenological 
considerations or through microscopic approaches. For example, 
Eq.~(\ref{eq:langevin}) can be obtained if one assumes that the excitations in 
the medium are small enough so that the medium is only slightly away from 
equilibrium and linear-response holds~\cite{Chandler, Caldeira-Leggett}. In a 
different approach, which we follow here, one uses an adiabatic perturbation 
theory which only assumes that $\boldsymbol{X}$ changes on a time scale that is 
longer than the relaxation time in
the medium. This approach has been applied successfully to both classical 
and quantum systems ~\cite{JAR,sivak_2012,berry_robbins,
Viola1,Viola2,zulkowski_2012,dalessio_2014}.

In what follows we use an adiabatic perturbation theory to consider a macroscopic object 
moving in a \textit{stochastic} environment comprised of many degrees of 
freedom. The stochastic environment evolves according to Markovian dynamics 
which obeys detailed balance so that, if the macroscopic object is stationary, the 
environment reaches thermal equilibrium with the probability of micro-state $s$ given 
by $P(s)=Z(\boldsymbol{X})^{-1}\, e^{-\beta H(\boldsymbol{X},s)}$.
Here $H(\boldsymbol{X},s)$ is the energy function of the environment,
$Z(\boldsymbol{X})$ is the partition function at fixed value of 
$\boldsymbol{X}$,
$\beta$ is the inverse temperature of the environment and we set the Boltzmann constant to one. 
Note that by using a Markovian environment we effectively assume that there are 
two time scales: i) a fast time scale associated with the transitions between 
microscopic configurations of the environment and ii) a longer time scale associated 
with the evolution of the macroscopic properties of the environment.  The 
adiabatic perturbation theory is carried out with respect to the second (slower) 
time scale.
For example, for a spin 
environment the first time scale is associated to individual spin flips while 
the second one is associated with the evolution of the total magnetization. The adiabatic perturbation theory assumes that the object moves slowly compared to the time scale associated to the evolution of the total magnetization so that the latter is almost equilibrated at every instantaneous position of the object.  

The organization of the paper is as follows. In Sec.~\ref{main} we introduce our main result for the friction coefficient and mass renormalization of the macroscopic object in contact with a dissipative stochastic environment. We then apply these results to a simple damped harmonic oscillator to show that, under appropriate conditions, the \textit{negative} mass correction can lead to an \textit{enhancement} of the oscillation frequency. In Sec.~\ref{simple} we study a more concrete model of a magnetic oscillator interacting with a spin environment in which our results can be derived
explicitly. By comparing the perturbative and the exact solution we show that an enhancement of the oscillation frequency can occur even when the adiabatic perturbation theory fails.
In Sec.~\ref{energy} we illustrate the consequence of the (negative) mass renormalization on the dynamics of the energy of the macroscopic object and show that, in general, the energy is not a monotonously decreasing function of time despite the fact that the object is in contact with a dissipative environment. 
Finally in Sec.~\ref{derivation} we present a detailed derivation of our results and systematically introduce the adiabatic perturbation theory.
Sec.~\ref{conclusion} contains our conclusions and future possible research directions.

\section{Main result\label{main}}

Our main result is that the first (irreversible and reversible) corrections to the motion of $\boldsymbol X$ from the adiabatic perturbation theory give rise to the following equations of motion:
\begin{equation}
\left(m+\kappa\right)\,\ddot{\boldsymbol{X}}+\eta\dot{\boldsymbol{X}}=-\frac{
\partial V}{\partial\boldsymbol{X}}\label{eq:langevin-renormalized} ~,
\end{equation}
after averaging over histories of the stochastic environment, where
\be
\begin{split}
\eta(t)&=\beta\int_{t}^{\infty}dt'\langle\partial_{X_{i}}H(t)\partial_{X_{i}}
H(t-t')\rangle_{0,c} ~~\mathrm{and}\\
\kappa(t)&=-\beta\int_{t}^{\infty}dt'\, 
t'\,\langle\partial_{X_{i}}H(t)\partial_{X_{i}}H(t-t')\rangle_{0,c}.\label{eq:eta_kappa}
\end{split}
\ee
are the friction coefficient and the mass correction respectively. 
Angular brackets indicate an average over the
equilibrium distribution of the medium at inverse temperature
$\beta$ at the instantaneous value of $X_{i}$ and $c$ indicates a connected correlation function. While
in general $\eta$ and $\kappa$ are tensors ($\eta_{i,j}$ and $\kappa_{i,j}$), we 
consider
the simple case where there is no preferred direction in space, such that
$\eta_{i,j}$ and $\kappa_{i,j}$ are proportional to the identity
matrix. Our perturbative treatment is systematic and can
be generalized to obtain higher order corrections in $\dot{\boldsymbol{X}}$
and, in relevant cases, the explicit tensorial behavior of $\eta$
and $\kappa$. The adiabatic perturbation theory carried out here assumes that: i) the environment relaxation time $\tau$ is much faster that the bare frequency of motion of the macroscopic object, i.e. $\omega_0 \tau \ll 1$, and that ii) during a relaxation time the object moves a distance smaller than $l$, where $l$ is the length over which $\tau$ and $\partial^2 V \over\partial X^2$ change, i.e. $\dot{\boldsymbol{X}}\tau\ll l$.
 When these two conditions are satisfied the dynamics of the macroscopic object is described by Eq.~\eqref{eq:langevin-renormalized} and the effect of the environment is contained in just two parameters, $\eta$ and $\kappa$. On the contrary when these conditions break down the dynamics of the object is not given by Eq.~\eqref{eq:langevin-renormalized} and one has to solve a full integral equation which describes both the action of the macroscopic object on the environment and the back-action of the environment on the macroscopic object. The detailed derivation of the adiabatic perturbation theory is given in Sec.~\ref{derivation}.
 
Eq.~\eqref{eq:eta_kappa} is identical to the high temperature limit of the 
expressions recently derived for quantum systems in Ref.~\cite{dalessio_2014}. 
However we note that in \cite{dalessio_2014} the derivation 
follows a different approach and makes different assumptions about the environment and,
 as we discuss later, leads to very different predictions. 
The expression for $\eta$ is well known from previous 
works~\cite{kirkwood,Hanggi}, for example, using linear response~\cite{kubo}.
It is always non-negative at positive temperatures. An important and interesting result, which to the best 
of our knowledge was previously overlooked, is that the mass correction 
$\kappa$ can be \textit{negative}. This, for example, happens when the dynamics of 
the bath is over-damped and the correlation function 
$\,\langle\partial_{X_{i}}H(t)\partial_{X_{i}}H(t-t')\rangle_{0,c}$ 
monotonically decreases in time. 
Both the mass correction and the dissipation originate from driving the medium out of equilibrium
through motion of the macroscopic particle. Both also depend on the coupling
between the object and the medium through $\partial_X H$.
When the coupling is weak, which is the region of main interest here, they are quadratic in the 
coupling strength, but in general their behavior can be more complicated.

\begin{figure}
\includegraphics[width=0.95\columnwidth]{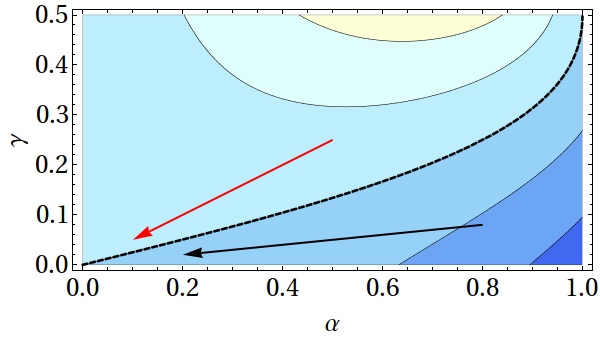}
\caption{(Color on-line) Contour plot of the ratio 
$\frac{\omega^{2}}{\omega_{0}^{2}}$ vs\add{.} $\gamma=\omega_0 \tau$ and $\alpha=\frac{\beta g^2 \tau}{m\omega_0}=\frac{\eta}{m \omega_0}$ 
where $\omega_0$ is the unperturbed frequency and $\tau=\frac{|\kappa|}{\eta}$. 
Along the dashed thick black line the correction due to friction and to the \textit{negative} mass renormalization balance each other and, despite being immersed in a dissipative medium, the object oscillates with its natural frequency, 
i.e. $\omega=\omega_0$. 
The red and black arrows represent the limit $\tau\rightarrow 0$, where perturbation
theory is best justified, for $\frac{\alpha}{\gamma}=\frac{\beta g^2}{m \omega_0^2}=2$ and $10$ respectively.  
The contour lines correspond to values from $0.8$ to $1.2$ (from blue to yellow) in steps of $0.1$.}
\label{fig:linear}
\end{figure}

A negative mass correction $\kappa$ has interesting consequences. First of all we note that any 
mass renormalization implies that the dynamics of the energy of the macroscopic object is non-Markovian (see Sec.~\ref{energy} below).
Moreover, for an oscillating macroscopic object a \textit{negative} mass correction can lead to an enhanced oscillation frequency in contrast to the usual suppression.
To describe this phenomenon we consider a damped harmonic oscillator, 
$V(\boldsymbol{X})=\frac{1}{2}m\,\omega_{0}^{2}\,\boldsymbol{X}^{2}$, in the case where the relaxation of 
the connected ``force-force" correlation function is given by:
\be
\langle\partial_{X}H(t)\partial_{X}H(0)\rangle_{0,c}=g^{2}e^{-t/\tau}\label{eq:two-point}
\ee
where $\tau$ is the relaxation time
of the medium and $g$ is the strength of 
a simple linear coupling between the macroscopic object 
and the medium. Substituting Eq.~\eqref{eq:two-point} into 
Eq.~\eqref{eq:eta_kappa}
we see that $\eta=\beta g^{2}\tau$ and $\kappa=-\beta g^{2}\tau^{2}$.
Therefore, when the adiabatic perturbation theory is justified, the emergent equation of motion is then given by:
\be
\left(m-\beta g^{2}\tau^{2}\right)\,\ddot{\boldsymbol{X}}+\beta 
g^{2}\tau\,\dot{\boldsymbol{X}}=-m\,\omega_{0}^{2}\,\boldsymbol{X}.
\label{eq:damped_osc}
\ee
Naively it might appear that $\kappa$ is higher order in $\tau$ than $\eta$ and 
therefore gives a negligible effect when the relaxation of the bath is fast 
(short $\tau$), but this conclusion can be misleading. It is easy to check 
that, to leading order, the shift in the frequency due to the mass correction and dissipation are:
\[
\delta\omega_{\kappa}^{2}\approx \omega_0^2\frac{\kappa}{m}=-\omega_0^2\frac{\beta g^{2}\tau^{2}}{m}, \quad \delta\omega_{\eta}^{2}\approx\frac{\eta^{2}}{4m^{2}}=\frac{\beta^{2}g^{4}\tau^{2}}{4m^{2}}.
\]
The two corrections have opposite sign and scale in the same way with $\tau$ so taking $\tau$ small 
does not guarantee that the frequency shift due to the dissipation dominates. 
The above expressions suggest that for small coupling between the system and the 
medium, or at high temperatures, the correction to the frequency due to the mass 
term $\delta\omega_{\kappa}$ dominates the frequency shift (see Fig.~\ref{fig:linear}). 
More generally it is easy to show that the ratio between the renormalized and unperturbed frequency 
is:
\[
\frac{\omega^{2}}{\omega_{0}^{2}}=\frac{1}{1- \gamma\alpha}\left(1-\frac{\alpha^2}{4(1-\gamma\alpha)}\right)
\]
where $\alpha=\frac{\beta g^2 \tau}{m\omega_0}=\frac{\eta}{m \omega_0}$ and $\gamma=\omega_0 \tau=\omega_0\frac{|\kappa|}{\eta}$.
For 
\be
0<\alpha<\frac{4 \gamma}{1+4\gamma^2} \label{eq:larger-condition}
\ee
the motion is under-damped and $\omega > \omega_0$. 
Specifically, in the small $\tau$ limit the frequency $\omega$ is enhanced 
if the ratio $\frac{\alpha}{\gamma}=\frac{\beta g^2}{m \omega_0^2}$, 
which is proportional to the square of the coupling strength, is smaller than $4$.
This confirms that the frequency is enhanced in the small coupling limit in agreement with our previous discussion.
We note that the validity of the adiabatic perturbation theory requires $\omega \tau\ll 1$. For $\omega>\omega_0$ this 
also implies $\gamma=\omega_0\tau\ll 1$. In addition, as explained in the Sec.~\ref{derivation}, we make the natural assumption that the properties of the environment, such as the the relaxation time $\tau$, change slowly with $\boldsymbol{X}$ so that they can be considered constant while the macroscopic object moves.
We also note that the condition $\gamma=\omega_0\tau\ll1$ together with $\alpha<4\gamma$ (see Eq.~\eqref{eq:larger-condition}) guarantees that the mass correction is small:
\[
\frac{\eta}{m\omega_0} <4 \omega_0\tau \rightarrow |\kappa| < 4 m (\omega_0\tau)^2 \ll m 
\]
where we have used $|\kappa|=\tau \eta$. Therefore when adiabatic perturbation theory applies and Eq.~\eqref{eq:langevin-renormalized} is valid, the total mass $m+\kappa$ is always positive (despite $\kappa$ being \textit{negative}).
Interestingly, the correction due to friction and to the \textit{negative} mass renormalization can balance each other and, despite being immersed in a dissipative medium, the object oscillates with its natural frequency (see Fig.~\ref{fig:linear}).

The above treatment relies on the adiabatic perturbation theory and 
simple assumptions about the behavior of the medium. 
Next we will study a more concrete model in which the results above can be derived explicitly.  In addition to demonstrating the 
results derived above in an explicit model,  
we show that the enhancement of the oscillation frequency can occur even when the 
adiabatic perturbation theory fails. Finally, this model will also highlight possible 
shifts in the oscillation frequency which may occur due to the interaction with 
the medium. These trivial corrections are akin to the Born-Oppenheimer terms in quantum mechanics 
and are expected to be absent when the oscillator is moving in a translationally 
invariant medium (see Sec.~\ref{derivation}).


\section{Magnetic oscillator in a spin environment\label{simple}}

\begin{figure}
\includegraphics[width=0.80\columnwidth]{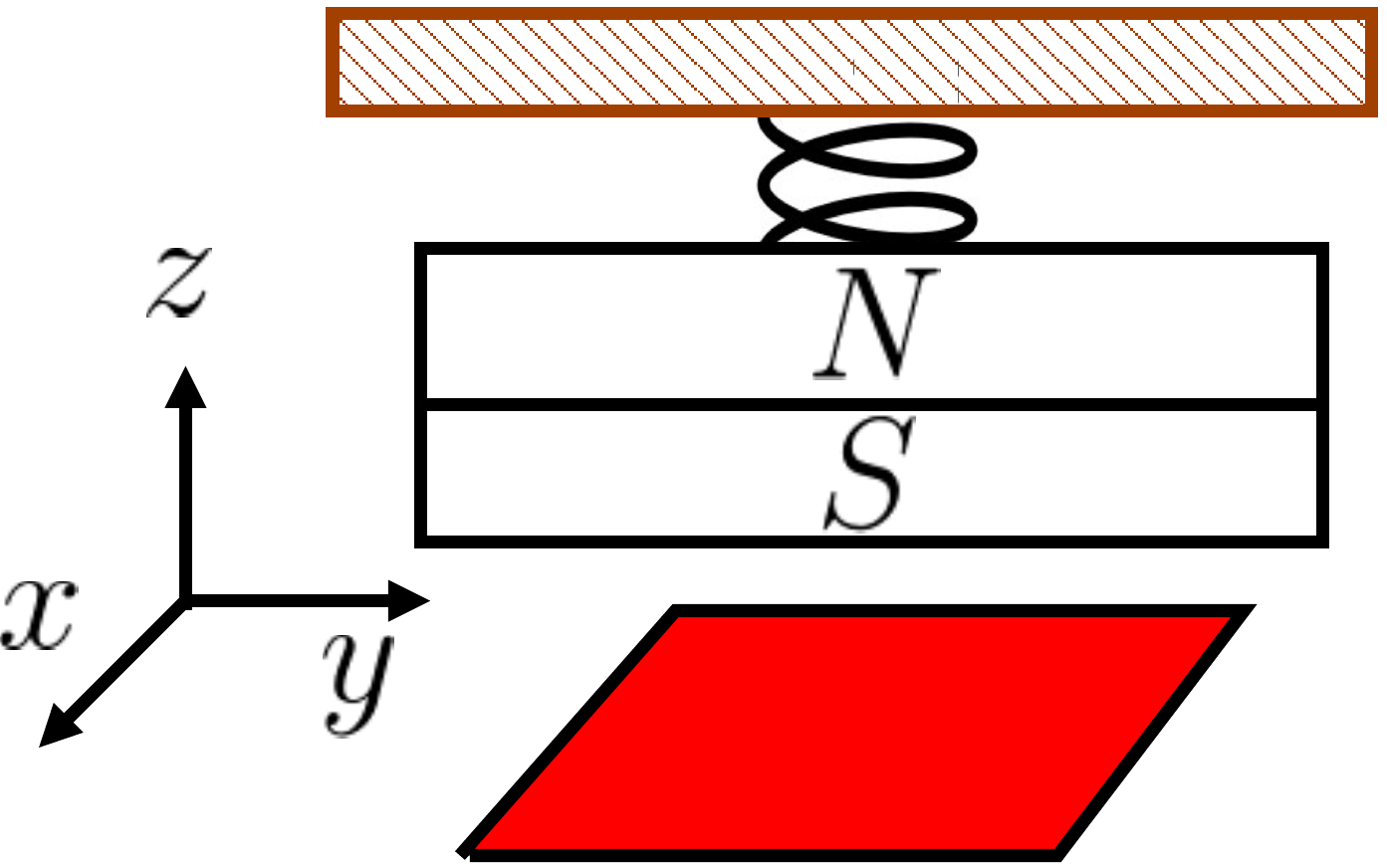}
\caption{(Color on-line) Schematic setup considered. A magnet oscillates
on top of a two dimensional spin system (indicated in red).}
\label{cartoon}
\end{figure}

Here we consider a magnetic oscillator which moves in the $z$ direction and which is coupled 
to a thin two-dimensional spin system which evolves under standard non-conserving spin dynamics (see Fig.~\ref{cartoon}).
The Hamiltonian of the oscillating magnet is:
\begin{equation}
H=\frac{p_z^2}{2m}+\frac{1}{2}m\omega_{s}^{2}\, z^{2}-h(z)\,\iint dxdy\, W(x,y)\,S(x,y)\label{eq:Hmagnet}
\end{equation}
where $S(x,y)$ is the (coarse grained) magnetization of the spin system at 
position $(x,y)$, $h(z)$ is the magnetic field at the spin plane which depends 
on the position $z$ of the magnet, $p_z$ is the momentum conjugate to $z$ and 
$\omega_{s}$ is the spring oscillation's frequency. 
The arbitrary weight function $W(x,y)$ describes the spatial variation of the magnetic field in the plane of the spins. When the weight function is flat, i.e. $W(x,y)=\text{const}$, the oscillator is coupled only to the zero Fourier component of the spin configuration while for a generic profile $W(x,y)$ the oscillator is coupled to many Fourier components of the spin configurations.
This Hamiltonian is valid when the thickness of the magnetic medium is negligible. 

The resulting equation of motion for the magnet is:
\be
\begin{split}
m\ddot{z} & \approx -m\omega_{s}^{2}\, z + h_1\,\iint dxdy\, W(x,y)\,S(x,y) \\
          & = -m\omega_{s}^{2}\, z+h_{1}\,\sum_{\boldsymbol{q}} S_{\boldsymbol{q}} W_{-\boldsymbol{q}} .
\end{split}
\label{eq:magnet-1}
\ee
where $S_{\boldsymbol{q}}$ and $W_{\boldsymbol{q}}$ are the Fourier components of the spin configuration and the profile function. In the equation above we have linearized the magnetic field near the equilibrium position: $h(z)\approx h_{0}+h_{1}\, z$.
This approximation is justified if the amplitude, $A$, of the motion of the magnetic oscillator is small. In particular we require that $h_1\gg h_2 A$ where $h_2\equiv \partial^2_{zz} h(z)|_{z=0}$. The effect of $h_2\ne0$ is discussed in Sec.~\ref{subleading}.

We model the spin environment via over-damped Langevin dynamics~\cite{kardar}:
\[
\partial_{t}S=-\mu\frac{\delta F}{\delta S}+\xi(x,y,t)\label{eq:langevin-spin}
\]
where $\mu$ is the mobility, $\xi$ is a white noise term which satisfies $\langle\xi\rangle=0$ and
\[
 \langle\xi(x,y,t)\xi(x',y',t')\rangle=2\frac{\mu}{\beta}\,\delta(x-x')\delta(y-y')\delta(t-t'),
\]
where $\beta$ is the inverse temperature of the spin system and the angular brackets denote an 
average over noise realizations.
$F$ is the Landau-Ginzburg free-energy:
\[
F=\iint dxdy\,\left[\frac{u}{2}|\nabla S|^{2}+\frac{r}{2}S^{2}-h(z)\, W(x,y)\,S(x,y)\right]
\]
where we have neglected higher order terms in $S$. We therefore obtain the following equation of motion for the Fourier components of the spin configuration:
\be
\partial_{t}S_{\boldsymbol{q}}=\mu\left[-(r+u |\boldsymbol{q}|^2) S_{\boldsymbol{q}} + (h_{0}+h_{1}z)W_{\boldsymbol{q}} \right] + \xi_{\boldsymbol{q}}(t) 
\label{eq:spin-1}
\ee
where $\xi_{\boldsymbol{q}}$ is the Fourier components of the noise. 

Next, we solve the coupled equations~\eqref{eq:magnet-1} and~\eqref{eq:spin-1}. First, for simplicity, we consider a case in which the weight function is flat so that the magnet is coupled only to the zero Fourier component of the spin configuration~\cite{lifted}. In particular in the equations above we set
$W_{\boldsymbol{q}=0}=1$ and $W_{\boldsymbol{q}\ne0}=0$ to obtain:
\be
\begin{split}
m\ddot{z}&=-m\omega_{s}^{2}\, z+h_{1}\, S_{0}\\
\partial_{t}S_{0}&=\mu\left(-r\, 
S_{0}+h_{0}+h_{1}z\right)+\xi_{0}(t)\label{eq:system-1}
\end{split}
\ee
It is then convenient to define $Z=\langle z\rangle-\langle z\rangle_{eq}$
and $\sigma_{0}=\langle S_{0}\rangle-\langle S_{0}\rangle_{eq}$ where
$\langle z\rangle_{eq}$,
$\langle S_{0}\rangle_{eq}$ are the stationary solutions of  $\langle z\rangle$ 
and $\langle S_{0}\rangle$ respectively:
\[
\langle z\rangle_{eq}=\frac{h_{0}h_{1}}{m\omega_{s}^{2}r-h_{1}^{2}},\quad\langle 
S_{0}\rangle_{eq}=\frac{m\omega_{s}^{2}h_{0}}{m\omega_{s}^{2}r-h_{1}^{2}}\;.
\]
Assuming that the system satisfies the stability condition  
$h_{1}^{2}<m r \omega_{s}^{2}$, the equations of motion reduce to: 
\begin{subequations}
\begin{align}
&&m\ddot{Z}=-m\omega_{s}^{2}\, Z+h_{1}\,\sigma_{0} \label{eq:system-2Aa}\\
&&\partial_{t}\sigma_{0}=\mu\left(h_{1}\, Z-r\,\sigma_{0}\right) 
\;.\label{eq:system-2A}
\end{align}
\label{systems}
\end{subequations}
These equations describe the motion of two degrees of freedom
with a linear coupling between them. If the relaxation of the spin system is fast 
($\mu r\gg \omega_s$)
the magnetization $\sigma_0$ follows closely the instantaneous equilibrium: 
$\sigma_0\approx h_1 Z/r$. 
This is a direct analogue of the Born-Oppenheimer approximation in quantum 
mechanics. 
Substituting $\sigma_0\approx h_1 Z/r$ into Eq.~(\ref{eq:system-2Aa}) we find a 
trivial shift of the oscillator frequency:
\be
\omega_0^2=\omega_s^2-{h_1^2\over mr} \label{softening}.
\ee
where $\omega_0$ includes the interaction effects with the environment 
and is therefore the direct analogue of the oscillation frequency in 
Eqs.~\eqref{eq:damped_osc}-\eqref{eq:larger-condition}.
Note that that stability condition guarantees that $\omega_0^2>0$. 

Before turning to the exact solution for $Z(t)$
we consider the adiabatic approximation. It is easy to show that the spin 
correlation function decays exponentially in time and it is given by 
$\langle S_0(0) S_0(-t') \rangle_{0,c} = e^{-t\mu r} / (\beta r)$~\cite{kardar}. 
The friction and the mass correction are therefore [see Eq.~\eqref{eq:eta_kappa}]:
\beq
&&\eta=\beta\, h_{1}^{2}\,\int_{0}^{\infty}dt'\,\langle 
S_0(0)S_0(-t')\rangle_{0,c}=\frac{h_{1}^{2}}{\mu r^{2}} \nonumber\\
&&\kappa=-\beta\, h_{1}^{2}\,\int_{0}^{\infty}dt'\, t'\,\langle 
S_0(0)S_0(-t')\rangle_{0,c}=-\frac{h_{1}^{2}}{\mu^{2}r^{3}}. \nonumber
\eeq
Therefore, the previous discussion holds if we identify 
\be
\tau=\frac{|\kappa|}{\eta}=\frac{1}{\mu r},\quad \beta g^2=\frac{h_1^2}{r},\quad \omega_0^2=\omega_s^2-{h_1^2\over mr}\label{eq:identify}.
\ee
Eq.~\eqref{eq:larger-condition} then gives the condition for an oscillation frequency larger than $\omega_{0}$.
Again we find that in the weak coupling regime, $\frac{h_{1}^{2}}{mr}\ll\omega_{s}^{2}$, the oscillation frequency is enhanced with respect $\omega_0$.
We note that within this linearized model it is impossible to obtain 
oscillation frequencies higher than the bare frequency $\omega_s$ because
the ``Born-Oppenheimer" softening (see Eq.~\eqref{softening}) is always larger than the 
frequency enhancement due to $\kappa$. 
However we stress that the ``Born-Oppenheimer" correction will 
be absent when the oscillator moves in a translationally 
invariant medium (see Sec.~\ref{derivation}) and, in that case, oscillation frequencies which are higher 
than the bare frequency $\omega_s$ can be achieved.

\begin{figure}
\includegraphics[width=0.95\columnwidth]{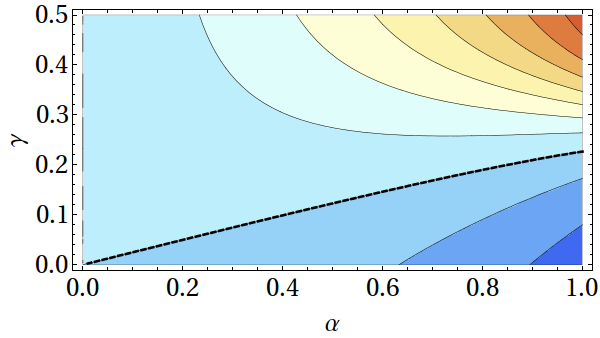}
\caption{(Color on-line) As in Fig.~(1) we show the contour plot of the ratio 
$\frac{\omega^{2}}{\omega_{0}^{2}}$ vs $\alpha$ and $\gamma$, where now the 
frequency $\omega$ corresponds to the imaginary part of the longest lived 
normal mode of Eq.~\eqref{eq:exact-1} (see text).
The dashed thick black line marks $\omega=\omega_0$.
Contour lines correspond to values from $0.8$ to $1.7$ (from blue to red) in steps of $0.1$.}
\label{fig:poles}
\end{figure}

To verify the validity of the adiabatic approximation in the interesting
regime where the frequency of oscillation is larger than $\omega_{0}$, 
we now solve the problem exactly. First we solve \eqref{eq:system-2A} to obtain
\[
\sigma_{0}(t)=e^{-\mu rt}\left[\sigma_{0}(0)+\mu h_{1}\int_{0}^{t}dt'\, e^{\mu rt'}Z(t')\right] \;.
\]
We now substitute this expression into Eq.~\eqref{eq:system-2Aa} to obtain
\[
\begin{split}
m\ddot{Z}(t)=&-m\omega_{s}^{2}\, Z(t)+h_{1}\sigma_{0}(0)\, e^{-\mu rt}\\
&+\mu h_{1}^{2}\int_{0}^{t}dt'\, e^{-\mu r(t-t')}Z(t') .
\end{split}
\]
Note that by expanding $Z(t')$ in the integral around $Z(t)$ the
approximate solution described above can be directly obtained.  
The above equation can be solved exactly using a Laplace transform, $\hat{Z}(s)=\int_{0}^{\infty}dt\, e^{-st}\, Z(t)$, to give:
\[
\hat{Z}(s)=\frac{m (\mu  r+s) (s Z(0)+\dot{Z}(0)) + h_1 \sigma_0(0)}{m (\mu  r+s) \left(s^2+\omega_s^2\right) - h_1^2 \mu}.
\]
Finally, we assume that the magnet and the spin system are initially
at equilibrium, $\sigma_{0}(0)=0$, $Z(0)=0$ and the magnet has an initial velocity $\dot{Z}(0)=v_{0}$ to obtain:
\begin{equation}
\hat{Z}(s)=v_{0} \left[ s^2+\omega_s^2-\frac{h_1^2}{mr}\frac{1}{(1+\frac{s}{\mu r})}\right]^{-1}.\label{eq:exact-1}
\end{equation}
The frequency of oscillations is given by the imaginary part of the poles of 
$\hat{Z}(s)$. Clearly, when $h_1^2/(mr)=0$, i.e. if the magnet and the spin system
are decoupled, the poles are at $s=\pm i\omega_{s}$ and free-oscillations are 
recovered. 
In Fig.~\ref{fig:poles} we show the value of the ratio $\omega^2/\omega_0^2$ 
as a function of $\alpha$ and $\gamma$ where, as before, $\alpha=\frac{\beta g^2 \tau}{m\omega_0}=\frac{\eta}{m \omega_0}$
and $\gamma=\omega_0\tau$ (see Eq.~\eqref{eq:identify}).
Here $\omega$ is given by the imaginary part of the solution of 
Eq.~\eqref{eq:exact-1} with the smallest real part, which corresponds to the slowest decaying mode. 
As expected in the limit $\alpha,\,\gamma\rightarrow 0$ the behavior 
of $\omega^2/\omega_0^2$ in Figs.~\ref{fig:linear} and 
\ref{fig:poles} are identical.
Note that, in this model, $\omega^2/\omega_0^2>1$ well 
beyond the limit where the adiabatic perturbation theory holds~\cite{foot1}.

\begin{figure}
\includegraphics[width=0.95\columnwidth]{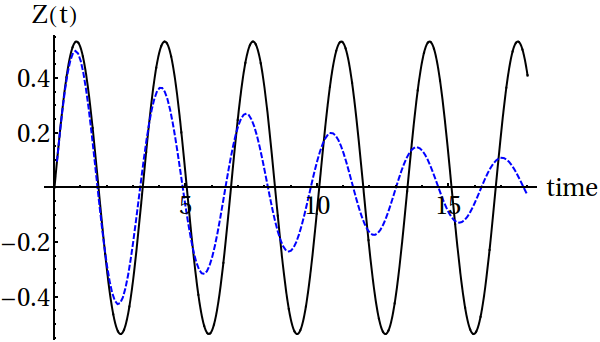}
\caption{(Color on-line) Comparison between the exact trajectory $Z(t)$ (dotted blue line) and the ``Born-Oppenheimer" approximation $Z_{BO}(t)=\frac{v_0}{\omega_0}\sin(\omega_0 t)$ (solid black line) for the magnetic oscillator interacting with many Fourier modes of the spin systems [see Eqs.~\eqref{rewriting} and \eqref{f(x)}]. The parameters are $u,r,h_1,m,\mu,v_0=1$, $R=0.01$, $L=2$ and $\omega_s=2.5$.
The value of $\omega_0\approx1.87$ is obtained from Eq.~\eqref{rewriting}.
We note that the exact solution oscillates at frequency $\omega>\omega_0$.}
\label{correction}
\end{figure}

The results found above for a simplified model, where the magnetic oscillator  
only couples to the $S_{0}\equiv S(\mathbf{q}=0)$ Fourier component of the magnetization, extend to more complex setups with only minor modifications. In particular, let us consider an arbitrary weight function $W(x,y)$ so that the magnet is coupled to many Fourier components of the spin configuration and show the the effect described above persists.
The previous procedure remains valid with minor modifications. 
For example Eq.~\eqref{systems} becomes:
\begin{subequations}
\begin{align}
m\ddot{Z}&=-m\omega_{s}^{2}\,Z+h_{1}\,\sum_{\boldsymbol{q}} \sigma_{\boldsymbol{q}} W_{-\boldsymbol{q}} \label{eq:system-2Bb}\\
\partial_{t}\sigma_{\boldsymbol{q}}&=\mu\left[ W_{\boldsymbol{q}} h_1 Z - (r + u |\boldsymbol{q}|^2) \sigma_{\boldsymbol{q}}\right]\;.\label{eq:system-2B}
\end{align}
\label{systemsB}
\end{subequations}
where, as before, $Z\equiv\langle z\rangle -\langle z\rangle_{eq}$ 
and $\sigma_{\boldsymbol{q}}\equiv \langle S_{\boldsymbol{q}}\rangle -\langle S_{\boldsymbol{q}}\rangle_{eq}$.
 However, this time the stationary solutions $\langle z\rangle_{eq}$ and $\langle S_{\boldsymbol{q}}\rangle_{eq}$ are:
\beq
&&\langle z \rangle_{eq} = \frac{h_0 h_1 K}{m\omega_s^2 r-h_1^2 K},  \nonumber\\
&&\langle S_{\boldsymbol{q}} \rangle_{eq}= \frac{h_0 W_q}{r + u |\boldsymbol{q}|^2} \frac{m\omega_s^2 r}{m\omega_s^2 r-h_1^2 K}\nonumber
\eeq
where $K\equiv\sum_{\boldsymbol{q}} \frac{|W_{\boldsymbol{q}}|^2}{1+(u/r) |\boldsymbol{q}|^2}$ and we have used that $W_{-\boldsymbol{q}}=W_{\boldsymbol{q}}^\star$. The equations~\eqref{systemsB} describe the effect of the magnetic oscillator on the spins and the back-action of the spins on the oscillator. Crucially, under our quadratic approximation for the Landau-Ginzburg free-energy $F$ the Fourier components of the spin configuration are decoupled. Moreover we have linearized the magnetic field around the equilibrium position of the oscillator, i.e. $h(z)\approx h_0+ h_1 z$. These two approximations allow to solve the above equation exactly for any profile $W(x,y)$. Repeating the same steps as before, Eqs.~\eqref{eq:exact-1} and~\eqref{softening} now become:
\beq
\hat{Z}(s)&&=v_0 \left[ (s^2+\omega_s^2)-\frac{h_1^2}{m r} f\left(s/(\mu r)\right) \right]^{-1} \nonumber \\
\omega_0^2&&=\omega_s^2-\frac{h_1^2}{m r}\,f(0) \label{rewriting}
\eeq
where we have defined the function: 
\be
f(x)\equiv \sum_{\boldsymbol{q}} \frac{|W_{\boldsymbol{q}}|^2}{x + 1 + (u/r) |\boldsymbol{q}|^2} \label{f(x)}.
\ee
When only the zero Fourier component contributes, i.e. $W_{\boldsymbol{q}=0}=1$ and $W_{\boldsymbol{q}\ne 0}=0$, we recover $f(x)=(1+x)^{-1}$ [see Eq.~\eqref{eq:exact-1}]. Here we consider the case in which many Fourier components contribute. For concreteness we assume that the interaction between the oscillator and the 
spin system has a Gaussian-like profile with width $R$ from which it follows that $|W_{\boldsymbol{q}}|^2=\exp\left[-(q_x^2+q_y^2)R^2\right]$ and \eqref{f(x)} 
becomes (we have replaced $\sum_{\boldsymbol{q}}\rightarrow \left(\frac{L}{2\pi}\right)^2 \int d\boldsymbol{q}$): 
\beq
f(x) = \left(\frac{L^2}{4\pi R^2}\right)\,y\,e^{y(x+1)}\,\Gamma[0,y(x+1)]
\eeq
where $y\equiv R^2 r/u$ and $\Gamma$ is the incomplete Gamma function. We note that in the limit of a broad profile, i.e. $y=R^2 r/u\gg1$, $f(x)\sim(1+x)^{-1}$ and we recover the case discussed for a flat weight function while in the narrow profile limit, i.e. $y\ll1$, $f(x)$ diverges logarithmically.  
We now substitute the expression above in Eq.~\eqref{rewriting} and perform the inverse Laplace Transform numerically to obtain the exact trajectory of the magnetic oscillator, $Z(t)$.
In Fig.~\ref{correction} we compare the exact trajectory $Z(t)$ with the ``Born-Oppenheimer" approximation $Z_{BO}(t)$ which describes oscillations at frequency $\omega_0$. For the parameters chosen the the mass correction is negative and the oscillation frequency is enhanced, i.e. $\omega>\omega_0$. This shows that the result obtained when the oscillator is coupled only to the zero Fourier component of the spin configuration persists even when the magnet is coupled to many degrees of freedom (many Fourier components) of the environment.

\section{Numerical simulations}

Let us now verify the results of the analytical calculations by direct numerical simulations of the microscopic equations of motion:
\[
m \ddot{z}(t) = -m \omega_{s}^{2}\, z(t)+h_{1}\,S_0(t).
\]
where $S_0(t)$ is the total spin magnetization and we set $m=1$ and $\omega_s=2.5$ (in what follows we use arbitrary units).
We introduce a time-discretization $\delta t=0.005$ and numerically compute the values $z_n\equiv z(n\, \delta t)$ using the velocity-Verlet algorithm~\cite{velocity-Verlet}. Between two subsequent updates (say $n$ and $n+1$), the position of the magnetic oscillator is assumed constant. In this time interval the spins are evolved according to Metropolis dynamics governed by the Hamiltonian:
\be
H_\text{Ising}=- \sum_{\la i,j\ra} s_i s_j -(h_0 + h_1 z_n) \sum_i s_i\,.
\label{H_ising}
\ee
Here $s_i$ indicates an Ising spin in a $2d$ system of linear size $L=350$ with periodic boundary conditions and the symbol $\la i,j\ra$ indicates a sum over nearest-neighbor. 
 A spin flip leads to a transition from configuration $C$ to a new configuration $C'$. This transition is determined by the rates $\Gamma(C\to C')$:
\[
\Gamma(C\to C') \equiv \frac{N}{\delta t}\, P(\Delta E) \equiv \frac{N}{\delta t}\, \text{min}\left(1,e^{-\beta \Delta E}\right)
\]
where $\beta$ is the inverse temperature of the spin system, $\Delta E$ is the energy difference between the two spin configurations and $N$ is the number of attempted spin flips during the a time interval $\delta t$. This choice of rates guarantees that the spin system relaxes toward the Gibbs equilibrium distribution at inverse temperature $\beta$ for any value of $z_n$. The value of $\beta=0.05$ is chosen such that the system is in the disordered phase ($\beta_c\approx 0.44$ for $h_0=h_1=0$). 

In practice, after each update of the position of the magnetic oscillator, we select at random $N$ spins, flip them according to probability $P(\Delta E)$ and compute the total magnetization $S_0$ which is then used to update the position of the magnetic oscillator. When $N$ is large the spin magnetization $S_0$ is able to relax to its equilibrium value, while when $N$ is small the spin magnetization ``lags behind" its equilibrium value. Therefore for large $N$ the motion of the magnetic oscillator is slow compared to the relaxation dynamics of the spin magnetization. This numerical scheme is similar to the one described in Ref.~\cite{magnetic-friction}.

In this setup the dynamics of the spin magnetization is expected to be described by the Langevin equation introduce in the previous section. However the parameters in the Langevin equation (such as $\mu$, $r$ and $u$) can not be derived rigorously from the microscopic rules introduced here. Therefore, to test our main result (namely that the oscillation frequency is enhanced with respect $\omega_0$) we need to compare three distinct simulations. In the first simulation we set  $h_0=h_1=0$ so that the spin system and magnetic oscillator are decoupled and, as expected, we obtain persistent oscillations of the magnetic oscillator at frequency $\omega_s$. In the second simulation, we set $h_0=h_1=7\,L^{-2}$ and $N=5\, L^2$. Now the magnetic oscillator is coupled to the total spin magnetization which is always fully relaxed to its (instantaneous) equilibrium value. Finally, in the third simulation, we keep $h_0=h_1=7\,L^{-2}$ but we set $N=0.01\,L^2$ so that the total spin magnetization is not fully relaxed to its (instantaneous) equilibrium value. Note that the couplings between the manetic oscillator and the spin systems are proportional to $L^{-2}$ while the total magnetization $S_0$ is proportional to $L^2$. This ensures that the equation of motion for the magnetic oscillator has a well defined thermodynamic limit. 

\begin{figure}
\includegraphics[width=1.0\columnwidth]{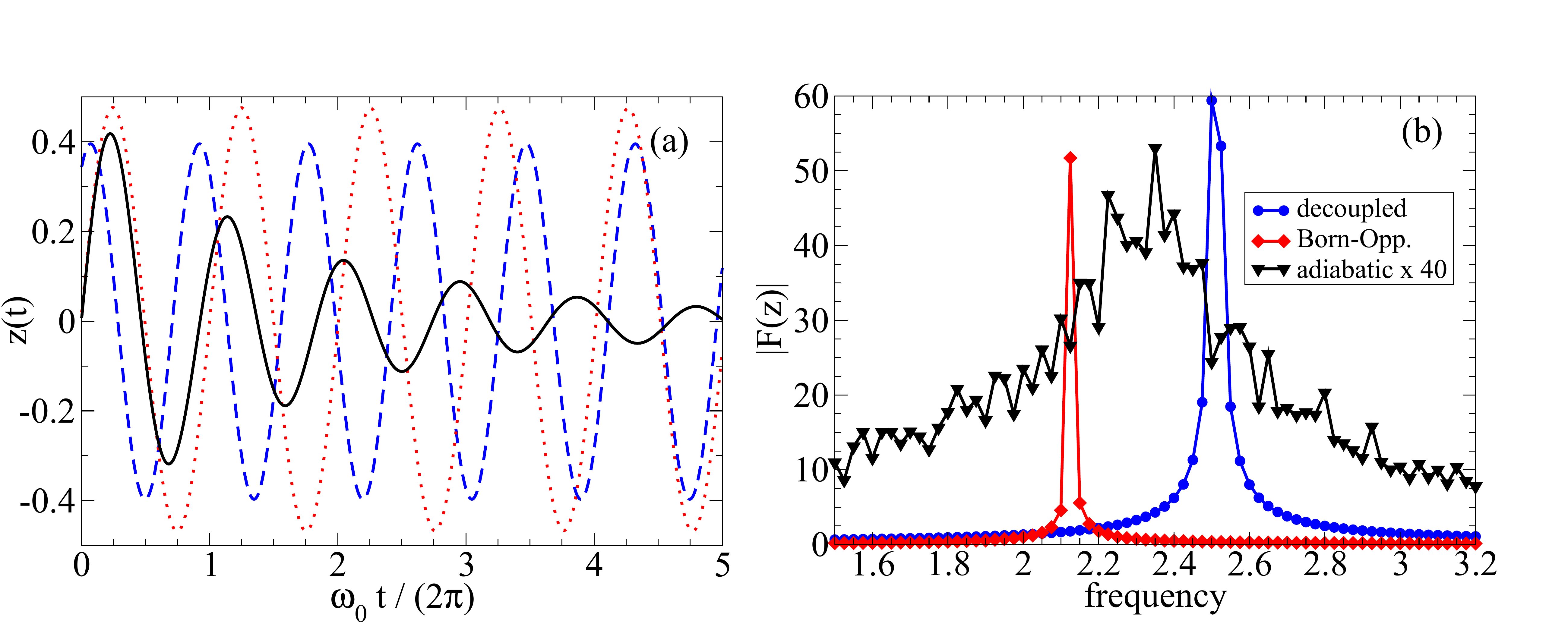}
\caption{(Color on-line) Three different microscopic simulations of the magnetic oscillator coupled to a spin system evolving according to Metropolis rates (see text for details). (a) The three trajectories $z(t)$ and (b) their Fourier components. In these simulations $\omega_s=2.5$, $\omega_0\approx 2.13$, $\omega\approx2.35$, $\alpha\approx 0.17$ and $\gamma=0.74$ where $\alpha,\,\gamma$ are defined as in Fig.~\ref{fig:linear} and Fig.~\ref{fig:poles}. Note that the black curve in panel (b) has been rescaled by a factor of $40$.}
\label{fig_numerics}
\end{figure}

The trajectories of the magnetic oscillator for the simulations described above are shown in Fig.~\ref{fig_numerics}(a). A Fourier analysis of these trajectories reveals important quantitative differences, see Fig.~\ref{fig_numerics}(b). In the first simulation only the frequency $\omega_s=2.5$ is present. In the second simulations, in which the magnet and the spin system are coupled and the spin magnetization is fully relaxed to its equilibrium value, the frequencies are narrowly peaked around a \textit{smaller} frequency which, according to the result of the previous section, can be interpreted as $\omega_0$ [see Eq.~\eqref{softening}]. In our case $\omega_0\approx 2.13$. In the third simulation, the frequencies are broadly peaked around at intermediate frequency $\omega_0<\omega<\omega_s$ with $\omega\approx2.35$ so that $\omega/\omega_0\approx 1.1$. The third simulation corresponds to the parameters $\alpha\approx 0.17$ and $\gamma=0.74$ where $\alpha,\,\gamma$ are defined in Fig.~\ref{fig:linear}. This confirms that, the leading correction above adiabaticity, lead to an \textit{enhanced} oscillation frequency with respect to $\omega_0$.

\section{energy absorption from the environment and negative mass\label{energy}}

To get some intuition for the negative mass correction, note that \textit{any} mass correction implies that the dynamics of the energy of the macroscopic object is non-Markovian. To see this we use Eq.~\ref{eq:langevin-renormalized} to write the time-derivative of its energy as:
\be
\frac{dE}{dt}=\frac{d}{dt}\left[ \frac{1}{2} m \dot{\boldsymbol{X}}^2 + V(\boldsymbol{X}) \right] = -\dot{\boldsymbol{X}}\left( \kappa \ddot{\boldsymbol{X}}+\eta \dot{\boldsymbol{X}} \right) 
\label{der_energy}
\ee
and note that it depends on $\ddot{\boldsymbol{X}}$.
The potential $V(\boldsymbol{X})$ includes the interaction energy with the environment and therefore $E(t)$ is conserved in the adiabatic limit. 
While the dissipative term ($-\eta \dot{\boldsymbol{X}}^2$) is always negative, the non-Markovian term  ($-\kappa \ddot{\boldsymbol{X}}\dot{\boldsymbol{X}}$) can change sign and become dominant close to turning points where $\dot{\boldsymbol{X}}=0$.
When $\kappa<0$ the macroscopic object transiently absorbs energy from the bath as it recedes from the turning points.
This holds when the bath is dissipative (or when its dissipative dynamics dominate over its inertial dynamics). In cases when $\kappa>0$ the system absorbs energy from the bath as it approaches the turning point. This is expected when the bath has a strong enough inertia so that, as the macroscopic object slows down at the turning points, the bath still ``pushes" it in the direction of the turning point. 

\begin{figure}
\includegraphics[width=0.95\columnwidth]{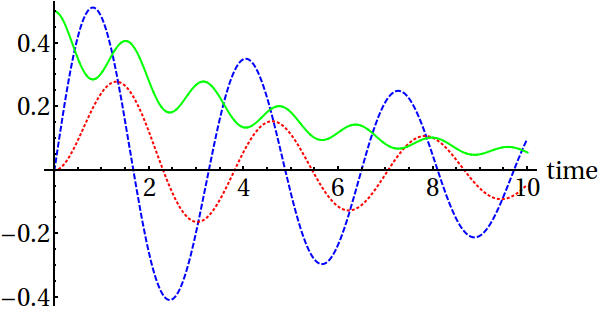}
\caption{(Color on-line) Exact time dependence of $Z(t)$ (dashed blue line) $\sigma_0(t)$ (dotted red line) and the energy (solid green line) of the magnetic oscillator $E(t)=\frac{1}{2}m\dot{Z}^2+\frac{1}{2}m \omega_s^2 Z^2 - \frac{h_1^2}{r} Z^2$.
The trajectories have been obtained by solving Eq.~\eqref{systems} exactly with the 
initial conditions $Z(0)=0,\,\sigma_0(0)=0$ and $\dot{Z}(0)=v_0=1$. 
The parameters $m,\,\mu,\,h_1$ are set to one while $r=1.1$ and $\omega_s=2$. 
These parameters correspond to $\tau\approx0.9$, $\omega_0\approx1.76$, $\gamma\approx1.59$ and $\alpha\approx0.47$.
In all the above we use arbitrary units. 
We note that for these parameters the mass correction is negative.}
\label{figSI1}
\end{figure}

To understand the fact that, when $\kappa<0$, the macroscopic object transiently absorbs energy from the bath as it approaches the turning point in Fig. \ref{figSI1} we show, on the same plot, the exact time evolution of $Z(t)$ and $\sigma_0(t)$ together with the instantaneous value of the equilibrium energy of the magnetic oscillator
\[
E(t)=\frac{1}{2}m \dot{Z}^2 + \frac{1}{2}m \omega_s^2 Z^2 - \frac{h_1^2}{r} Z^2 .
\]
The first two terms are the kinetic and elastic potential energies while the last term represent the ``Born-Oppenheimer" interaction potential energy.
The last term is obtained by substituting in the interaction energy, $U_{int}=- h_1 \sigma_0 Z$, $\sigma_0$ with 
its equilibrium value $\sigma_0^{eq}=\frac{h_1}{r}Z$ (see Eq.~\eqref{eq:system-2A}). 

The trajectories for $Z(t)$ and $\sigma_0(t)$ are obtained by solving Eqs.~\eqref{systems} numerically. We observe that, for a short time after the object leaves the turning point, the magnet absorbs energy from the environment. This is due to the fact that away from the adiabatic limit $\sigma_0$ is not in equilibrium at the instantaneous value of $Z(t)$ but instead it ``lags'' behind its equilibrium value. When the system approaches the $Z>0$ turning point $\sigma_0<\sigma_0^{eq}\equiv\frac{h_1 Z}{r}$. 
and the environment applies a force (proportional to the lag) which decreases $|\dot{Z}|$. At the turning point, $Z$ changes direction 
but $\sigma_0$ is still smaller than $\sigma_0^{eq}$. This force is now increasing $|\dot{Z}|$ which causes the object to transiently absorb energy from the bath. 
To see this more explicitly we combine Eqs.~\eqref{eq:system-2Aa} and \eqref{eq:system-2A} to write the equation of motion for the macroscopic object as: 
\[
m\ddot{Z}(t)=-m\omega_{0}^{2}\, Z -\frac{h_{1}}{\mu r} \partial_t\sigma_{0}
\] 
This shows explicitly that an additional force, proportional to $\partial_t \sigma_0$, acts on the magnet because the magnetization $\sigma_0$ has not reached equilibrium at the instantaneous value of $Z$. If this was the case, $\partial_t\sigma_0$ would be zero and the equation above would simply describe oscillations at frequency $\omega_0$.
Consider the object moving towards the positive $Z$ turning point. As it approaches the turning point $\sigma_0$ trails behind $Z$ so that $\partial_t \sigma_0$ is positive (see Eq. \eqref{eq:system-2A}). The force experienced by the magnet pulls it away from the turning point, slowing it down. Right after the turning point, since $\sigma_0$ still trails behind $Z$,  $\partial_t \sigma_0$ is still positive pulling it away from the turning point causing it to absorb energy from the environment. This chain of events can be clearly seen in Fig. \ref{figSI1}.

We stress that Eq.~\eqref{der_energy} only holds in the long time limit when $\eta$ and $\kappa$ have reached their asymptotic values (see Eq.~\eqref{eq:eta_kappa}). In particular if the environment is initially (at $t=0$) in equilibrium then $E(t)<E(0)$ for any $t>0$ and any initial condition of the oscillator (including $\dot{Z}(0)=0$ and $Z(0)$ at a turning point). This is required by the second law of thermodynamics. However, we note that neither the second law of thermodynamics nor the condition $E(t)<E(0)$ imply that the function $E(t)$ decreases monotonically. In fact the opposite is true as shown in Fig.~\ref{figSI1}.

The initial decrease of the energy can be understood from the short time behavior of the force acting on the macroscopic object which is given by (see Eq.~\eqref{force} below)
\[
\begin{split}
F(t=0) = &- \langle \frac{\partial H}{\partial X}\rangle = -\langle \frac{\partial H}{\partial X} \rangle_0 \\
& - \beta \dot{X}(t) \langle \partial_{X} H(t=0)\partial_{X} H(t=0) \rangle_{0,c} dt.
\end{split}
\]
where, for simplicity, we have assumed that the macroscopic object is described by a single degree of freedom and, as in Eq.~\eqref{eq:eta_kappa}, the suffix ``$0$" indicates that the averages are over the
equilibrium distribution of the environment at inverse temperature
$\beta$ and $c$ indicates a connected correlation function.
This implies that, to leading order in $dt$, the energy change (including the trivial ``Born-Oppenheimer" term) is given by: 
\[
\begin{split}
\frac{dE}{dt} &=F(0)\dot{X}(0) \\
& =-\beta \dot{X}^2(0) \langle 0| \partial_{X} H(0)\partial_{X} H(0)| 0 \rangle_c <0
\end{split}
\]
which is always negative (for an initial state with positive temperature). The statement that, if the environment is initially (at $t=0$) in equilibrium then $E(t)<E(0)$ can be made rigorous for any $t>0$~\cite{inpreparation}.

\begin{figure}
\includegraphics[width=0.95\columnwidth]{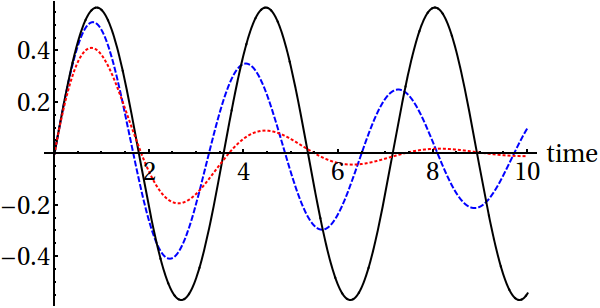}
\caption{(Color on-line) Comparison between the exact trajectory $Z(t)$ (dotted blue line), 
the ``Born-Oppenheimer" approximation $Z_{BO}(t)=\frac{v_0}{\omega_0}\sin(\omega_0 t)$ (solid black line) and the ``Born-Oppenheimer" approximation with the inclusion of dissipation $Z_{\eta}$ (dotted red line) obtained by solving the equation $m \ddot{Z}=-m\omega_0^2 Z-\eta Z$. The parameters are as in Fig.~\ref{figSI1}. On one hand, $Z_{\eta}(t)$ oscillates at a reduced frequency, $\omega<\omega_0$, and quickly decay to zero as expected for a damping ratio $\zeta=\eta/\left(2 m \omega_{0}\right)\approx 0.23$. On the other hand, the exact solution oscillates at frequency $\omega>\omega_0$ and decay to zero more slowly. When $\gamma=\omega_0\tau\ll1$ the difference between these trajectories decreases.}
\label{figSI2}
\end{figure}

In Fig.~\ref{figSI1} the parameters are chosen so that the mass correction is negative and the oscillation frequency is enhanced, i.e. $\omega>\omega_0$. This can be seen clearly in Fig.~\ref{figSI2} where we compare the exact trajectory $Z(t)$ with the ``Born-Oppenheimer" approximation $Z_{BO}(t)$ which describes oscillations at frequency $\omega_0$ and $Z_{\eta}$ which includes both the ``Born-Oppenheimer" correction and the effect of the dissipation. We note that $Z_{\eta}$ oscillates, as expected, at a reduced frequency $\omega<\omega_0$. We stress that the two approximations, $Z_{BO}(t)$ and $Z_{\eta}(t)$, become asymptotically exact in the adiabatic limit. Finally, we stress that when the inertia dominates the dynamics of the environment we expect the mass correction to be positive. 

\section{Adiabatic Perturbation Theory for Markov Processes\label{derivation}}

Finally, in this Section we present a detailed derivation of the main results Eqs.~\eqref{eq:eta_kappa}.
We consider a Markov process whose rates change as a function of a set of external 
parameters ${\bf X}(t)$ which evolve in time. We are interested in developing an 
adiabatic perturbation theory in the rate of change of these external parameters. 
To this end, start with the master equation
\begin{equation}
\partial_t | P(t) \rangle ={\bf M}({\bf X}(t))| P(t) \rangle 
\label{master}
\end{equation}
where ${\bf M}({\bf X}(t) )$ is a Markov matrix. The vector $|P\rangle$  
specifies the probability of being in a given configuration. The off 
diagonal elements of the matrix ${\bf M}$ specify transition rates and are 
positive. The diagonal elements are set by conservation of probability, 
$M_{ii}=-\sum_{j\neq i} M_{ji}$. The matrix ${\bf M}$ is in general not symmetric 
and therefore admits separate left and right eigenvectors (for more details see \cite{vanKampen,tauber}). 
It is easy to see that there is a trivial left eigenvector $\langle 0|$ 
with eigenvalue zero and all entries equal to one:
\be
\langle 0|=\langle 1,1,\dots, 1|,\; \langle 0| {\bf M}=0 \label{M_0}.
\ee
With these definitions the average of a given observable, ${\cal O}$ is given 
by:
\begin{equation}
\langle {\cal O} \rangle = \langle 0 | \mathcal O| P \rangle=\sum_{ij} \mathcal 
O_{ij} P_j
\end{equation}
The diagonal values of the operator $\mathcal O$ specify the values of the 
observable $\mathcal O$ in  configuration $|j\rangle$. Off diagonal terms are 
can also occur, for example when $\mathcal O$ measures currents. 

Next, we define the adiabatic basis of the instantaneous eigenvectors of ${\bf M}({\bf X}(t))$:
\be
{\bf M}({\bf X}(t))|n({\bf X}(t)) \rangle=\epsilon_n({\bf X}(t))|n({\bf X}(t)) \rangle 
\label{basis}
\ee
where the normalization of the vectors is such that 
$\langle n({\bf X} ) | m({\bf X}) \rangle = \delta_{n,m}$. 
$\epsilon_n({\bf X}(t))$ is the eigenvalue at a fixed
value of of ${\bf X}(t)$.
Eigenvalues of Markov matrices are known to have obey $\text{Re}\left[\epsilon_n({\bf X}(t))\right]\leq 0$ \cite{vanKampen,tauber}. 
We now rewrite the master equation in the instantaneous basis, see Eq.~\eqref{basis}. (This step is analogue to using a moving reference frame in classical mechanics.) 
Using 
\begin{equation}
 | P(t) \rangle = \sum_n a_n(t)|n({\bf X}(t)) \rangle \;,
\end{equation}
in Eq.~(\ref{master}) gives:
\begin{eqnarray}
\sum_n \left( \dot{a}_n |n({\bf X}) \rangle + a_n  \frac{\partial}{\partial t}|n({\bf X} )\rangle \right)=\sum_n a_n 
\epsilon_n({\bf X})  |n({\bf X}) \rangle\;.\nonumber
\label{moving_frame1}
\end{eqnarray}
where, if necessary, the time-derivative of $|n({\bf X} )\rangle$ can be evaluated using the chain rule:
\[
 \frac{\partial}{\partial t}|n({\bf X} )\rangle  \equiv \dot{X}_i \frac{\partial}{\partial {X_i}} |n({\bf X} )\rangle\,.
\]
To simplify notations we have suppressed, for now, the explicit dependence of the 
quantities on time. We will reintroduce the time-dependence explicitly when necessary to avoid confusion.
Projecting on $\langle m({\bf X})|$ gives the set of coupled equations:
\begin{equation}
\dot{a}_m=\epsilon_m({\bf X}) a_m -\sum_n \langle m({\bf X}) | \partial_t |n({\bf X} )\rangle  a_n \;. \label{eq:master2}
\end{equation}
Next, it is useful to define
\begin{equation}
a_m(t) = \tilde{a}_m(t)\exp\left[\int_0^t \epsilon_m(t)dt'\right] \;,
\end{equation}
where the dynamics start at time $t=0$. Then using Eq.~(\ref{eq:master2}) we have
\begin{equation}
\dot{\tilde{a}}_m(t)=-\sum_n \tilde a_n(t) e^{\int_0^t 
(\epsilon_n(t')-\epsilon_m(t'))dt'} \langle m({\bf X}) | \partial_t |n({\bf X} )\rangle \nonumber,
\end{equation}
which in turn is equivalent to the integral equation
\beq
\tilde{a}_m(t)=-\int_0^t dt' \left[\sum_n \tilde a_n(t')  e^{\int_0^{t'} (\epsilon_n(t'')-\epsilon_m(t''))dt''}  \right. \nonumber \\
\biggl. \langle m({\bf X}(t')) | \partial_{t'} |n({\bf X}(t') )\rangle \biggr]  
+{\tilde{a}}_m(0)\;.
\eeq
In terms of the original $a_m$ we then have 
\begin{multline}
a_m(t)=e^{\int_{0}^t \epsilon_m(t')dt'}a_m(0)\\
-\int_0^t dt'\sum_n a_n(t')  e^{\int_{t'}^{t}\epsilon_m(t'')dt''}
\langle m({\bf X}(t')) | \partial_{t'} |n({\bf X} (t'))\rangle \label{formal}.
\end{multline}
Recall that Markov matrices always have an eigenstate $|0({\bf X}) \rangle$ with 
zero eigenvalue, i.e. $\epsilon_0({\bf X})=0$ \cite{vanKampen,tauber}. This eigenvector corresponds to 
the steady-state of the system at fixed ${\bf X}$. For this state we then have
\beq
a_0(t)&&=-\int_0^t dt' \sum_n a_n  (t') \langle 0({\bf X}(t')) | \partial_{t'} |n({\bf X}(t'))\rangle+a_0(0) \nonumber\\
&&=a_0(0)=1. \label{prob_cons} 
\eeq
The equality follows from differentiating the orthogonality relation
\[
\partial_{t} \langle 0({\bf X}(t))|n({\bf X}(t))\rangle=0
\]
and noting that 
\[
\partial_{t} \langle 0({\bf X}(t))|=\partial_{t}\langle 1,1,\dots 1|=0.
\]
so that 
\[
\langle 0({\bf X}(t)) | \partial_t |n({\bf X}(t) )\rangle=0\; 
\forall n.
\]
Equation~\eqref{prob_cons} simply reflects the conservation of probability.
Up to this point all the expressions are exact. 
In particular Eq.~\eqref{formal} is a rewriting, in a particular (co-moving) basis, of the master equation Eq.~\eqref{master}. We now develop a formal perturbation theory in the time derivatives of ${\bf X}$. 
This is possible because, by assumption, the eigenvalues $\epsilon_m$ and eigenvectors $|n\rangle$ depends on time only through ${\bf X}(t)$.
For example the energies which appear in Eq.~\eqref{formal} are:
\be
\int_{t'}^{t}\epsilon_m(t'')dt''\equiv \int_{0}^{t-t'} \epsilon_m(t-q) dq \label{en} 
\ee
Expanding in a Taylor series in $q$ and using the chain rule $\partial_t =\dot{{\bf X}}\frac{\partial}{\partial{\bf X}}$ we arrive at:
\beq
\epsilon_m(t-q)&&\approx \epsilon_m(t)-q \frac{\partial\epsilon_m}{\partial t} +\dots \nonumber\\
&&\approx \epsilon_m({\bf X}(t))-q \left.\left(\dot{{\bf X}}\frac{\partial\epsilon_m}{\partial{\bf X}}\right)\right|_t+\dots \nonumber
\eeq
Substituting back into Eq.~\eqref{en} we have:
\be
\int_{0}^{\delta t} \epsilon_m(t-q) dq \approx  \epsilon_m(t)\delta t -\frac{\delta t^2}{2} \left(\dot{{\bf X}}\frac{\partial\epsilon_m}{\partial{\bf X}}\right)+\dots \label{en_series}
\ee
where we have defined $\delta t\equiv t-t'$. This expression shows that if
\[
\gamma_1=\frac{\delta t \dot{{\bf X}}}{\epsilon_m} \frac{\partial\epsilon_m}{\partial{\bf X}} \ll 1 
\]
then only the first order on the RHS of Eq.~\eqref{en_series} contributes. 
The dimensionless parameter $\gamma_1$ describes the relative change of the eigenvalues $\epsilon_m$ during a time-interval $\delta t$. Specifically it measures the change in relaxation time during a time $\delta t$. This is naturally expected to be small, in particular when the coupling to the bath is weak. This dimensionless parameter can be made arbitrary small by decreasing the velocity $\dot{{\bf X}}$ and we can therefore conclude that at small velocities:
\[
\int_{t'}^{t}\epsilon_m(t'')dt''\approx \epsilon_m(t) (t-t') 
\]
where, importantly, the energies are evaluated at the final time $t$.
A similar argument applies to the matrix elements in Eq.~\eqref{formal}. 
In fact the relative change of the matrix element during a time interval $\delta t$  
\[
\gamma_2\equiv  \delta t \dot{X}_i \frac{ \frac{\partial}{\partial X_i} \left(\langle m({\bf X}) |  \frac{\partial}{\partial X_j} |n({\bf X})\rangle\right)}{\langle m({\bf X}) |  \frac{\partial}{\partial X_j} |n({\bf X})\rangle}
\]
can be made arbitrary small by reducing the velocity $\dot{X}_i$. 
Therefore, at sufficiently small velocity, $\gamma_1,\gamma_2\ll1$ and both the energies and the matrix elements can be considered constant and can be evaluates at the final time $t$. Then Eq.~\eqref{formal} reduces to:
\beq
a_m(t)=&&e^{\epsilon_m(t) t}a_m(0)-\int_0^t dt'\sum_n a_n(t') \times \nonumber\\
 &&e^{\epsilon_m(t) (t-t')} \dot{X}_i(t') \langle m({\bf X}(t)) | \frac{\partial}{\partial X_i} |n({\bf X}(t))\rangle \nonumber.
\eeq
We now assume that we start at the steady-state of the system at ${\bf X}(t=0)$ so that $a_m$ with $m \neq 0$ are small quantities. 
Then, it is easy to see that the lowest order contribution to the amplitude $a_m$ is linear in the velocity and originates from the term $n=0$ (the contribution from $a_n$ with $n\ne0$ is, at least, of order $\dot{X}^2$):
\be
a_m(t)=-\int_0^t dt'e^{\epsilon_m(t)(t-t')} \dot{X}_i (t') \langle m({\bf X}(t)) |\frac{\partial}{\partial {X_i}} |0({\bf X}(t) )\rangle \label{formal2}
\ee
where we used $a_0=1$.

The above derivations apply to any Markov process. We now restrict 
ourselves to processes which begin in equilibrium and whose dynamics satisfy 
detailed balance for any fixed value of ${\bf X}$.
We stress that the detailed balance condition constraint the eigenvalues of the Markov matrix to be real.
Then the steady-state eigenvector is given by
\begin{equation}
|0({\bf X}) \rangle = {1\over Z({\bf X})}\sum_s e^{-\beta H({\bf X},s)}|s 
\rangle \;,
\label{O(X)}
\end{equation}
where $|s \rangle$ is a vector with a $1$ entry at configuration $s$ and zero 
otherwise, $H({\bf X},s)$ is the energy of state $s$ at a given value of ${\bf 
X}$ and $\beta$ is the inverse temperature. $Z$ is the partition function of the 
model at a given value of ${\bf X}$. Then it is straightforward to see that
\begin{multline}
\langle  m({\bf X})|\frac{\partial}{\partial X_i}|0({\bf X}) \rangle =  \\
 -{1\over Z} \sum_s \langle m({\bf X})|\left( \partial_{X_i} (\beta H)-\langle 
\partial_{X_i} (\beta H) \rangle\right)e^{-\beta H({\bf X},s)} | s \rangle.   \label{multi}
\end{multline}
Here the angular brackets denote an average with respect to the equilibrium 
measure at a given value of ${\bf X}$. Next, note that  from the orthogonality 
relation $\langle m ({\bf X})|0({\bf X})\rangle=0$ for $m\neq 0$ the second term in the relation 
above drops out, so that
\begin{eqnarray}
\langle m({\bf X})|\frac{\partial}{\partial X_i}|0({\bf X}) \rangle &=& -{1\over 
Z}\sum_s  \langle m({\bf X})|\partial_{X_i} (\beta H)e^{-\beta H({\bf X},s)} |s \rangle 
\nonumber \\
&=&  -  \langle m({\bf X})|\partial_{X_i} (\beta H)| 0({\bf X}) \rangle.
\label{derivative}
\end{eqnarray}
Therefore
\begin{eqnarray}
a_m(t)= \int_0^t dt'  e^{\epsilon_m(t)(t-t')}  \dot{X}_i (t')\langle m({\bf X}(t)) | 
\partial_{X_i}(\beta H) |0({\bf X}(t))\rangle \nonumber
\label{eq:at3}
\end{eqnarray}

In the paper our interest is in calculating thermodynamic generalized forces 
$\langle \partial_{X_j} (\beta H) \rangle$.  While in the paper we assume that 
$\beta$ is constant, here we also allow for it to serve as an externally varying
parameter (see for example~\cite{zulkowski_2012}). Using
\begin{equation}
|P(t)\rangle = |0 ({\bf X}(t))\rangle +\sum_{m \neq 0} a_m(t) |m ({\bf X}(t))\rangle 
\end{equation}
we have
\begin{multline}
\langle 0 | \partial_{X_j} (\beta H)|P ({\bf X}(t))\rangle = \langle 0 | \partial_{X_j} (\beta H)|0 ({\bf X}(t))\rangle \\
+ \int_0^t dt'  e^{\epsilon_m(t)(t-t')} \dot{X}_i (t') \cdot \\
\sum_{m \neq 0} \langle 0| \partial_{X_j} (\beta H)|m ({\bf X}(t))\rangle \langle m({\bf X}(t)) | \partial_{X_i}(\beta H) |0({\bf X}(t))\rangle.
\end{multline}
where we have used that the left zero eigenvalue in independent on ${\bf X}(t)$ (see Eq.~\eqref{M_0}).
Using the relation $e^{\epsilon_m(t)(t-t')} |m \rangle = e^{{\bf M}(t)(t-t')}|m \rangle$ this can be rewritten as:
\begin{multline}
\langle 0 | \partial_{X_j} (\beta H)|P ({\bf X}(t))\rangle = \langle 0 | \partial_{X_j} 
(\beta H)|0 ({\bf X}(t))\rangle \\
+ \int^t_{0}dt'\dot{X}_i(t') \sum_{m\ne0}\langle 0| \partial_{X_j} (\beta H) e^{{\bf M}(t)(t-t')} |m ({\bf X}(t))\rangle \\
\times \langle m({\bf X}(t)) |\partial_{X_i} (\beta H)| 0({\bf X}(t))\rangle = \langle \partial_{X_j} (\beta H)(t) \rangle_0\\
+ \int^t_{0}dt'\dot{X}_i(t') \langle \partial_{X_j} (\beta H)(t') \partial_{X_i} (\beta H)(t)\rangle_{0,c} \\
= \langle\partial_{X_j}(\beta H)(t)\rangle_0 + 
\int^t_{0}dt'\dot{X}_i(t-t')\,\mathcal{K}_{0,c}(t-t',t)
\label{force}
\end{multline}
where the subscript $0$ indicates that the expectation value are evaluated 
in the instantaneous equilibrium at fixed ${\bf X}(t)$ and $\mathcal{K}_{0,c}$ is a short-hand notation 
for the two-times force-force connected correlation function evaluated at equilibrium.
In the long time limit ($t$ much longer than the relaxation time $\tau$ of $\mathcal{K}_{0,c}$) the integral 
above is dominated by short $t'\le \tau$ implying that one can extend the upper 
limit of integration to $\infty$. Note that the existence of a finite $\tau$  
essentially assumes that the spectrum of the Markov matrix is gapped. Finally we have
\begin{multline}
\langle \partial_{X_j} (\beta H) (t)\rangle = \langle\partial_{X_j}(\beta H)(t)\rangle_0 \\
+ \int^\infty_{0}dt'\dot{X}_i(t-t')\,\mathcal{K}_{0,c}(t-t',t)
\label{eq:pert}
\end{multline}

For a particle moving in a stochastic environments the equations of motion (averaged over histories of the stochastic environment) are given by
\begin{equation}
m\,\ddot{\boldsymbol{X}}=-\left\langle \frac{\partial H}{\partial\boldsymbol{X}}\right\rangle\;.
\end{equation} 
Using the above results and keeping $\beta$ fixed we have
\begin{equation}
\begin{split}
m\,\ddot{\boldsymbol{X}}=&-\langle \partial_{X_j}  H (t) \rangle_0 \\
&- \beta \int^\infty_{0}dt'\dot{X}_i(t-t') \mathcal{K}_{0,c}(t-t',t) \\
=& F_\text{BO} -\eta \dot{\boldsymbol{X}} - \kappa \ddot{\boldsymbol{X}} + O(\dddot{\boldsymbol X})~,
\end{split}
\end{equation}
where we have identified the ``Born-Oppenheimer" contribution to the force 
\[
F_\text{BO}=-\frac{\partial V}{\partial {\bf X}}=-\langle \partial_{{\bf X}} H(t)\rangle_0=-\langle 0 |\partial_{{\bf X}} H|0({\bf X}(t))\rangle
\]
and expanded $\dot{X}_i(t-t')$ around $t'=0$ to obtain $\eta$ and $\kappa$ defined as in Eq.~\eqref{eq:eta_kappa}. 
The ``Born-Oppenheimer" force can be easily computed from the following identity [see Eq.~\eqref{O(X)}]:
\[
\partial_{\bf X} \langle 0 | 0({\bf X})\rangle = - \langle 0 | \partial_{\bf X}(\beta H)| 0({\bf X})\rangle - \frac{1}{Z({\bf X})}\frac{\partial Z({\bf X})}{\partial {\bf X}} \langle 0 | 0({\bf X})\rangle
\]
where $Z({\bf X})$ is the partition function at fixed value of ${\bf X}$. Using the orthogonality relation $\langle 0 | 0({\bf X})\rangle=1$ for any ${\bf X}$ we obtain:
\be
F_\text{BO}=\frac{1}{\beta Z({\bf X})} \frac{\partial Z({\bf X})}{\partial {\bf X}} \label{FBO}
\ee
which shows that the ``Born-Oppenheimer" force depends exclusively on the energy spectrum and vanish if $Z({\bf X})=\text{const.}$. This is the case, for example, when the unperturbed system is translationally invariant.

\subsection{Subleading corrections in the small system-bath coupling limit\label{subleading}}

We now compute the first subleading correction in the limit of small
system-bath coupling. We stress that our adiabatic perturbation theory does not rely on the small system-bath coupling
however, in this limit, it is easy to explicitly compute the subleading corrections (which
are non-linear in the velocity) and show that they are indeed negligible at small velocities.

We start from the exact equation Eq.~\eqref{formal} and assume that the system is at steady-state so that $a_{m}$ with
$m\ne0$ are small quantities. It is easy to see that the lowest non-vanishing
contribution to $a_{m}$ is of first order in the system-bath coupling.
This contribution is obtained by substituting in Eq.~\eqref{formal} the
right eigenvectors computed to first order in the system-bath coupling
and the unperturbed eigenvalues and left eigenvectors. Noting that
the unperturbed eigenvalues and eigenvectors are independent on $X$
and are therefore time-independent we arrive at: 
\[
a_{m}(t)=-\int_{0}^{t}dt'e^{\epsilon_{m}(t-t')}\dot{X}(t')\langle m|\frac{\partial}{\partial X}|0(X(t'))\rangle
\]
where we used $a_{0}=1$ (the contributions from $a_{n}$ with $n\ne0$
are higher order in the system-bath coupling).

We now apply the expression:
\[
f\left(X(t')\right)=f\left(X(t-t'')\right)=f\left(X\right)-t''\dot{X}\frac{\partial f}{\partial X}+\dots,\quad
\]
where $t''\equiv t-t'$ and the RHS is evaluated at $t$ to obtain:
\begin{multline}
a_{m}(t)=-\int_{0}^{t}dt''e^{\epsilon_{m}t''}\dot{X}(t-t'')\times \\
\left[ \langle m|\frac{\partial}{\partial X}|0(X)\rangle-t''\dot{X}(t)\,\langle m|\frac{\partial^{2}}{\partial X^{2}}|0(X)\rangle+\dots\right] 
\end{multline}
Using the form of the steady state eigenvector [see Eq.~\eqref{O(X)}] and manipulations 
similar to the one in Eqs.~\eqref{O(X)}--\eqref{derivative} we obtain that to lowest order in the system-bath coupling:
\begin{multline}
\begin{split}
\langle m|\frac{\partial}{\partial X}|0(X)\rangle &= -\langle m|\frac{\partial}{\partial X}(\beta H)(t)|0\rangle \\
\langle m|\frac{\partial^{2}}{\partial X^{2}}|0(X)\rangle &=-\langle m|\frac{\partial^{2}}{\partial X^{2}}(\beta H)(t)|0\rangle
\end{split}
\end{multline}
where $|0\rangle$ is the unperturbed stationary state. Then we compute
\[
\langle0|\partial_{X}(\beta H)|P(X)\rangle,\quad|P(X)\rangle=|0(X)\rangle+\sum_{m\ne0}|m(X)\rangle
\]
which, to lowest order in the system-bath coupling, gives [we have followed the same manipulations as in Eq.~\eqref{force}]:
\begin{multline}
\langle0|\partial_{X}(\beta H)|P\rangle=\langle0|\partial_{X}(\beta H)|0\rangle+\int_{0}^{\infty}dt''\dot{X}(t-t'')\times\\
\Bigl[\langle0|\partial_{X}(\beta H)(t-t'')\partial_{X}(\beta H)(t)|0\rangle\Bigr. \\
\left.-\dot{X}(t)\, t''\,\langle0|\partial_{X}(\beta H)(t-t'')\partial_{XX}^{2}(\beta H)(t)|0\rangle\right] \label{formal2}
\end{multline}
where we have extended the upper limit of integration to infinity by noting that this integral is dominated by $t''\lesssim \tau$ (where $\tau$ is the correlation time).

In the example of the magnetic oscillator considered above the interaction energy between the
macroscopic degrees of freedom and the environment is: 
\[
H_{int}=-h(X)\, S_{0}\approx-\left(h_{0}+h_{1}X+h_{2}X^{2}+\dots\right)S_{0}
\]
where $S_{0}$ is the zero Fourier component of the magnetization
and we have expanded the magnetic field $h(X)$ to second order around
the equilibrium position (in Sec.~\ref{simple} we considered only the
first order). We now see that both correlation functions appearing in Eq.~\eqref{formal} are proportional to $\langle0|S_{0}(t-t'')S_{0}(t)|0\rangle=\frac{e^{-t''/\tau}}{\beta r}$.
More importantly by expanding $\dot{X}(t-t'')\approx\dot{X}(t)-t''\ddot{X}(t)$ in Eq.~\eqref{formal2} we can explicitly read the coefficients which multiply $\dot{X}$, $\ddot{X}$ and $\dot{X}\dot{X}$.
These coefficients are respectively the friction coefficient $\eta$, mass renormalization $\kappa$ and the non-linear coefficient $\eta_2$:
\[
\eta=\frac{\beta h_1^2}{r}\,\tau,\quad \kappa=-\frac{\beta h_1^2}{r}\,\tau^2,\quad \eta_2=-\frac{\beta h_1 h_2}{r}\,\tau^2.
\]
As expected we reproduce the expressions for $\eta$ and $\kappa$ in the Sec.~\ref{simple} [recall that $\tau=1/(\mu r)$]. 
We can now explicitly see when the quadratic contribution in the velocity is negligible compared to the friction and the mass renormalization:
\beq
 &&|\dot{X}\dot{X}\eta_2| \ll |\eta \dot{X}|    \rightarrow (\omega_0 \tau) A/l \ll 1 \nonumber \\ 
 &&|\dot{X}\dot{X}\eta_2| \ll |\kappa \ddot{X}| \rightarrow A/l \ll 1 
\label{condition}
\eeq
where we have used that for an oscillating object $\dot{X}\sim \omega_0 A$, $\ddot{X}\sim \omega_0^2 A$ and $l\equiv h_1/h_2$ is the characteristic length over which the system bath coupling changes.
The two conditions in Eq.~\eqref{condition} can be satisfied when both dimensionless parameters $\omega_0 \tau$ and $A/l$ are small.
The first condition simply states that the system is slow compared to the relaxation time of the environment. 
Naively, the second condition, $A\ll l$, limits the validity of our results to the small amplitude oscillation regime. 
This is misleading since $A$ should be understood as (at most) the distance covered during a single relaxation time, i.e. $A\sim \dot{X}\tau$. 
This is obvious if one realizes that the environment relaxes to equilibrium on the time-scale $\tau$ and therefore the dynamics of the system can not depend on phenomena which occurred too far in the past. Therefore the condition $A\ll l$ can be satisfied (for any $\tau$) at sufficiently small velocity signifying that the expressions derived in the Sec.~\ref{simple} become asymptotically exact in the limit of small velocity. In particular, for the magnetic oscillator example, the quadratic contribution to the velocity can be neglected for $\dot{X}\ll v_\text{cr}$ where:
\be
v_\text{cr} \sim \frac{l}{\tau} = \frac{\mu r h_{1}}{h_{2}}.
\ee


\section{conclusion\label{conclusion}}

In conclusion we have studied the dynamics of a macroscopic object coupled to
a stochastic environment using an adiabatic perturbation theory. The
dynamics of the stochastic environment are taken to obey detailed
balance. Surprisingly we find that the emergent equations of motion
contain, besides the usual dissipative term, a \textit{negative} mass
correction. Naively this term is higher order in perturbation theory
than the dissipative term and therefore negligible. However, as we
demonstrated, this can be misleading. For example, the frequency
shift in a damped harmonic oscillator resulting from both terms
is of the same order (in $\tau$). 

Moreover, the dissipative term breaks time-reversal
symmetry while the mass correction does not. This means that these two
quantities capture fundamentally different properties and should
be analyzed separately. This is exemplified in Sec.~\ref{energy} where we have analyzed the dynamics of the energy of the macroscopic object. While dissipation would lead to a monotonic energy decrease the mass correction leads to the non-Markovian dynamics. In fact the macroscopic object can transiently absorb energy from the environment and its energy is, in general, a non-monotonically decreasing function of time despite the fact that the object is coupled to a dissipative environment.

Correction to the mass in emergent equations of motion are by no means
uncommon. For example in hydrodynamics it is well known (see for 
example~\cite{landau}) that
interaction of, say, a ball with a fluid lead to both a dissipative
term and a mass correction. However, these mass corrections are always positive
in contrast to the \textit{negative} mass correction discussed above.
Interestingly, when the coupling between the environment and the object is via the momentum it is possible to find a negative mass correction for a particle in the strong coupling regime~\cite{momentum2,momentum3}. However, that setup is rather different from the situation described here where we focus on real space interactions and the negative mass correction appears in the weak coupling limit.

Remarkably, within a similar adiabatic perturbation theory in quantum
systems the mass correction can be shown to be strictly positive~\cite{dalessio_2014}.
It remains unclear how the two approaches can be made to agree and
what this implies either on the adiabatic perturbation theory in quantum
mechanics or the validity of stochastic dynamics for classical systems with many 
degrees of freedom. We plan to address this question in a future publication.

Finally, we note that the above treatment has been carried out to an order which is
formally identical to a linear-response regime~\cite{kubo}. It will be 
interesting to see if
higher order terms can lead to other interesting effects.

\textit{Acknowledgments.} We are grateful to M. Kolodrubetz for the careful reading of the manuscript and helpful comments.
This work was partially supported by BSF 2010318 (YK and AP), NSF
DMR-1506340 (LD and AP), AFOSR FA9550-10-1-0110 (LD and AP), ARO W911NF1410540 (LD and AP) and ISF grant (YK).
LD acknowledges the office of Naval Research.
YK is grateful to the BU visitors program.


\begin{thebibliography}{References}
\bibitem{lifted} This assumption is lifted later and we show that the effect described persists even when the oscillator is coupled to many Fourier components.
\bibitem{foot1} We also note that $\omega<\omega_s$, meaning that the frequency of oscillation 
is always reduced compared to the bare oscillation frequency of the spring.
\end{thebibliography}
\end{document}